\documentclass[aps,pre,twocolumn]{revtex4-2}
\usepackage{graphicx} % Required for inserting images
\usepackage{xcolor}
\usepackage{hyperref}
\usepackage[normalem]{ulem} % For strikethrough with \sout
\usepackage{amsmath}

\begin{document}
\title{Comment on ``Role of Matter Interactions in Superradiant Phenomena''}

\author{Max Hörmann}
% \email{max.hoermann@fau.de}

\author{Anja Langheld}
\email{anja.langheld@fau.de}

\author{Jonas Leibig}
% \email{jonas.leibig@fau.de}

\author{Andreas Schellenberger}
% \email{andreas.schellenberger@fau.de}

\author{Kai Phillip Schmidt}
\email{kai.phillip.schmidt@fau.de}

\affiliation{Friedrich-Alexander-Universit\"at Erlangen-N\"urnberg, Department of Physics, Staudtstraße 7, 91058 Erlangen, Germany}

\begin{abstract}
    Recently, Mendonça \textit{et al.}~\cite{Mendonca2025} investigated the Dicke-XXZ model and the Dicke-Ising model. 
    For the latter model, their calculated quantum phase diagram contradicts claims about the existence of an intermediate phase with superradiant and antiferromagnetic order and the change in order of some phase transition lines, observed in other studies \cite{Zhang2014,Rohn2020,Langheld2024a,RomanRoche2025}. 
    In this comment we demonstrate that both features are indeed present in the Dicke-Ising model for the investigated parameter range in  Ref.~\cite{Mendonca2025}.
\end{abstract}

\maketitle

\section{Introduction}

In their recent article, Mendonça \textit{et al.}~\cite{Mendonca2025} present a novel numerical approach to tackle one-dimensional composite light-matter systems.
By using a variational unitary transformation accompanied by density matrix renormalization group (DMRG) calculations, they map out the quantum phase diagram of Dicke-spin models.
For the case of the Dicke-Ising model, which is introduced in Eq.~(1) of Ref.~\cite{Mendonca2025} as
\begin{align*}
        \mathcal H &= \omega a^\dagger a + \varepsilon \sum_{i=1}^N s_i^z + \frac {2g}{\sqrt N} \sum_{i=1}^N s_i^x (a+a^\dagger) \\&\phantom{=}- 4J \sum_{\langle i,j\rangle } s_i^{z} s_j^{z}
\end{align*}
with spin-1/2 operators $s_i^{\alpha}$, some of their findings contradict other studies performed on this model \cite{Langheld2024a, RomanRoche2025, Rohn2020, Zhang2014}, some of which were argued to be limited by their methods \cite{Zhang2014,Rohn2020}, while others were not even mentioned \cite{Langheld2024a, RomanRoche2025}.
This comment presents strong evidence for why Ref.~\cite{Mendonca2025} misses two key features of the quantum phase diagram, namely,
\begin{enumerate}
    \item the existence of an intermediate antiferromagnetic superradiant phase and
    \item the change in order of phase transitions for ferromagnetic Ising interactions.
\end{enumerate}
In the following sections we will discuss these issues separately, followed by conclusions.

\begin{figure}
    \centering
    \includegraphics[width=\linewidth]{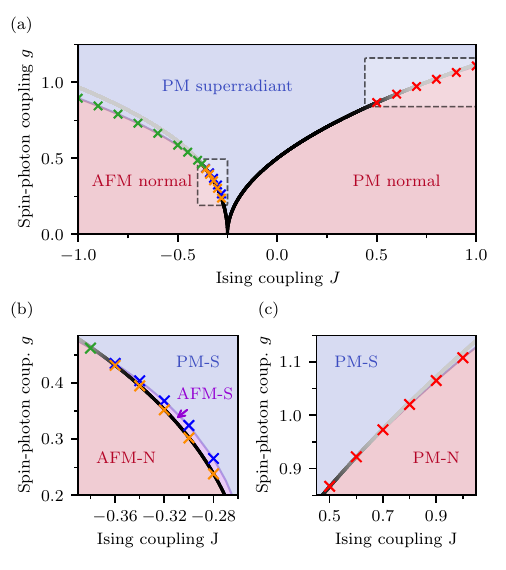}
    \caption{ 
        Phase diagram of the Dicke-Ising model, choosing $\omega=\varepsilon=1$ and similar parameter ranges for $g,J$ as in Fig.~3(c) from  Ref.~\cite{Mendonca2025}. 
        Crosses and background colors show the QMC results using the algorithm from Ref.~\cite{Langheld2024a}, black lines depict the mean-field results \cite{Zhang2014,Schellenberger2024}.
        The black mean-field lines fade when they start to deviate from the QMC analysis. 
        (a) Full phase diagram containing antiferromagnetic normal (AFM-N), antiferromagnetic superradiant (AFM-S), paramagnetic normal (PM-N), and paramagnetic  superradiant (PM-S) phases, (b) Zoom into intermediate AFM-S phase, which was not found by Ref.~\cite{Mendonca2025}, (c) Zoom into emerging deviation from mean-field results \cite{Zhang2014, Schellenberger2024} for ferromagnetic Ising interactions not included in Ref.~\cite{Mendonca2025}.
        }
    \label{fig:phase_diagram}
\end{figure}

\section{On the intermediate antiferromagnetic superradiant phase}

For the case of antiferromagnetic (AFM) interactions, mean-field calculations observe an intermediate phase with coexisting antiferromagnetic order and superradiance \cite{Zhang2014}.
The existence of this intermediate phase is quantitatively verified by the quantum Monte Carlo (QMC) simulations of Langheld \textit{et al.}~\cite{Langheld2024a} in one and two dimensions with system sizes up to $N=8192$ spins.
As indicated by the orange and blue crosses in Fig.~\ref{fig:phase_diagram}, the intermediate phase is of quite limited extent in one dimension (compared to higher dimensions \cite{Langheld2024a}).
While we do not have further insight into the analysis of Ref.~\cite{Mendonca2025}, the data points shown in the figures of their work indicate that the resolution of their calculations is of similar order as the extent of the intermediate phase. 
This makes it difficult to find this phase, especially considering their relatively small system size of presumably up to $N=100$ for the Dicke-Ising model.
The quantum phase transition from the normal (non-superradiant) to the intermediate phase is of second order, as motivated by the exact solution of the low-energy spectrum in the antiferromagnetic normal phase by Ref.~\cite{Schellenberger2024} and verified by QMC \cite{Langheld2024a}.

\section{On the order of phase transitions for ferromagnetic Ising interactions}

In the realm of ferromagnetic interactions in the Dicke-Ising chain, the authors of Ref.~\cite{Mendonca2025} claim to find a second-order phase transition to the superradiant phase.
However, going to the limit of a vanishing longitudinal magnetic field $\epsilon$, the system gains an additional spin-flip symmetry \cite{Rohn2020}, forcing the quantum phase transition to be of first order by Landau theory.
This first-order transition point is also determined numerically with arbitrary precision using an exact self-consistent approach (see Appendix of Ref.~\cite{Langheld2024a}), which has been confirmed by exact analytical calculations \cite{RomanRoche2025}.
Both approaches exploit the mapping to an effective Hamiltonian that becomes exact in the thermodynamic limit. 
The phase transitions remain first order for sufficiently small longitudinal magnetic fields, as confirmed by QMC \cite{Langheld2024a}.
For larger longitudinal fields, the order of the quantum phase transition changes at the multi-critical point at \mbox{$J\approx \varepsilon / 2 = 0.5$ \cite{Langheld2024a}} (see Fig.~\ref{fig:phase_diagram}). 
Thus, the first-order phase transition line lies outside of the parameter regime discussed in Ref.~\cite{Mendonca2025}, making it plausible that it could also be tracked when going to larger Ising interactions in Fig.~\ref{fig:phase_diagram}.

\section{Conclusions}

In this comment we stress the existence of an intermediate phase with magnetic and superradiant order in the antiferromagnetic Dicke-Ising chain, as well as the change in the order of the quantum phase transition in the ferromagnetic Dicke-Ising chain, which were both missed in the discussion by Mendonça \textit{et al.}~\cite{Mendonca2025}.
For these features we rely on quantitative numerical calculations done on the full light-matter system at large system sizes \cite{Langheld2024a} and on analytical findings that hold exact in the thermodynamic limit (see Appendix of \cite{Langheld2024a} and \cite{Schellenberger2024,RomanRoche2025}).
To further check on the novel numerical approach by \cite{Mendonca2025}, it would be interesting to investigate if their approach is able to find these features, when going to larger system sizes, higher resolution, and a larger parameter space.

\vspace{0.5cm} % Fixing broken layout

\begin{acknowledgments}

The authors gratefully acknowledge the support by the Deutsche Forschungsgemeinschaft (DFG, German Research Foundation) - Project-ID 429529648 - TRR 306 \mbox{QuCoLiMa} (``Quantum Cooperativity of Light and Matter") and the Munich Quantum Valley, which is supported by the Bavarian state government with funds from the Hightech Agenda Bayern Plus. We further gratefully acknowledge the scientific support and HPC resources provided by the Erlangen National High Performance Computing Center (NHR@FAU) of the Friedrich-Alexander-Universit\"at Erlangen-N\"urnberg (FAU). The hardware is funded by the German Research Foundation (DFG).

\subsection*{Data availability}
The QMC data and the points of phase transition extracted from it for this work are openly available \cite{RawData}.

\subsection*{Author contributions}
MH: Conceptualization, Writing -- review \& editing.
AL: Conceptualization, Formal analysis, Data curation, Investigation, Visualization, Software, Writing -- review \& editing.
JL: Conceptualization, Writing -- review \& editing, Validation.
AS: Conceptualization, Formal analysis, Writing -- original draft, Writing -- review \& editing.
KPS: Conceptualization, Supervision, Writing -- review \& editing \footnote{Following the taxonomy \href{https://credit.niso.org}{CRediT} to categorize the contributions of the authors.}.
\end{acknowledgments}

\bibliography{bibliography.bib}

\end{document}